\documentclass{cimento}

\usepackage{amsmath,amssymb}
\usepackage{dsfont}
\usepackage{bm}
\usepackage{tikz}
\usetikzlibrary{calc,matrix,arrows,decorations.pathmorphing,decorations.markings}

\newcommand{\bvec}[1]{{\ensuremath{\mathbf{\bm{#1}} }}}
\newcommand{\Dif}{\text{d}}
\newcommand{\braket}[3]{\ensuremath{\left< #1 \vphantom{#3} \right| #2 \left| #3 \vphantom{#1} \right>}}

\title{Evolution and dynamics of cusped light-like Wilson loops}
\author{F.F.~Van~der~Veken\from{ins:ua}}
\instlist{
	\inst{ins:ua}
	Departement Fysica, Universiteit Antwerpen, B-2020 Antwerpen, Belgium\\
	\hspace*{0.5em}frederik.vanderveken@ua.ac.be
}
\PACSes{
	\PACSit{11.10.Gh}{Renormalization}
	\PACSit{11.15.Tk}{Other nonperturbative techniques}
	\PACSit{11.25.Tq}{Gauge/string duality}
}

\begin{document}

\maketitle

\begin{abstract}
We discuss the possible relation between the singular structure of TMDs on the light-cone and the geometrical behaviour of rectangular Wilson loops.
\end{abstract}

\section{Introduction}
The singularity structure of transverse momentum dependent parton density functions (or TMDs for short) is known to be a lot more complex than that of collinear parton density functions. For the most common singularities this doesn't really pose a problem (\emph{e.g.} ultraviolet poles can be removed by general methods, like standard renormalisation using the $R$-operation). However, light-like singularities can be much more difficult to treat. A TMD is considered `light-like' when at least one of its segments is on-light-cone (on-LC for short). In this case it is not entirely clear whether standard renormalisation remains a sufficient technique, due to the emergence of extra overlapping divergences. A standard TMD can be defined as \cite{TMDdef}:
\begin{equation}
	f(x,\bvec{k}_\perp) = 
		\frac{1}{2}\int \frac{\Dif z^-\Dif^2 \bvec{z}_\perp}{2\pi(2\pi)^2} \; e^{ik\cdot z}
		\braket{P,S}{\bar{\psi}(z) \, U^\dag (z;\infty) \gamma^+ U(\infty;0) \, \psi(0)}{P,S} \Big|_{z^+=0}
\end{equation}
where the Wilson lines are split into their longitudinal and transversal parts:
\begin{align}
	U(\infty, 0) &=
		U(\infty^-,\bvec{\infty}_\perp ; \infty^-, \bvec{0}_\perp) U(\infty^-,\bvec{0}_\perp ; 0^-, \bvec{0}_\perp)\\
	&=
		\mathcal{P} \,\text{exp}\left[-i g \int_{0}^{\infty} \Dif z_\perp \; A_\perp(\infty^-, \bvec{z}_\perp)\right]
		\mathcal{P} \,\text{exp}\left[-i g \int_{0}^{\infty} \Dif z^- \; A^+(z^-, \bvec{0}_\perp)\right]
		\nonumber
\end{align}
The overlapping divergences, coming from the light-like behaviour of the TMD, will manifest themselves as terms of the order $1/\epsilon^2$ or, depending on the regulator used in the gluon propagator, $ \left(\ln\epsilon\right) / \epsilon$, where $\epsilon$ is the regulator used in dimensional regularisation. These overlapping divergences will give the only contribution to the evolution equations, which are governed by the cusp anomalous dimension. The latter is given by \cite{cad1,cad2}
\begin{equation}
\label{eq:cad}
	\Gamma_{\text{cusp}} = \frac{\alpha_s \, C_F}{\pi} \left( \chi \coth \chi - 1 \right)
	\quad \stackrel{\text{on-LC}}{\longrightarrow} \quad
	\frac{\alpha_s \, C_F}{\pi}
\end{equation}
where $\chi$ is the cusp angle. In the on-LC limit we have $\chi\rightarrow\infty$, which renders the cusp anomalous dimension independent on $\chi$ (in fact, it becomes infinite, but on-LC it will be redefined as the factor in front of the double pole). For this reason it is common in literature to refer to this setup as having a `hidden cusp'.

In what follows we will take a closer look on Wilson loops, which are Wilson lines on a closed path. Although that at a first glance they seem totally uncorrelated to TMDs, there might exist a profound relation between both. We will try to show that a Wilson loop on a specific path, namely a rectangular loop on the null plane and with light-like segments, has a singularity structure which can be related to that of on-LC TMDs. This is an important motivation for the possible existence of a duality between those two objects.

\section{Wilson loops and loop space}
As mentioned before, a Wilson loop is a Wilson line on a closed path. Because a Wilson loop should be a fully self-contained object by definition, all indices (Lorentz and Dirac) are traced over, and the loop is evaluated in the ground state. Then a general Wilson loop can be written as
\begin{equation}\label{eq:wloop}
	\mathcal{W}[C] =
		\frac{1}{N_c}\,\text{tr}\, \braket{0}{\mathcal{P} \,\text{exp}\left[i g \oint_C \Dif z^\mu A^a_\mu (z) t_a \right]}{0}
\end{equation}
where $C$ is any closed path and $A_\mu^a$ is the (non-Abelian) gauge field, evaluated in the fundamental representation. This loop is a pure phase, transforming coordinate dependence into path dependence. As is known (see \cite{QCDrecast1,QCDrecast2}), Wilson loops can be used as elementary building bricks to completely recast QCD in loop space. To achieve this, the definition of a Wilson loop needs to be extended to make it (possibly) dependent on multiple contours. We define a $n$-th order Wilson loop (consisting of $n$ sub-loops) as
\begin{equation}
	\mathcal{W}_n(C_1,\ldots, C_n) =
		\braket{0}{\Phi(C_1) \ldots \Phi(C_n)}{0}
\end{equation}
where each sub-loop is defined as
\begin{equation}
	\Phi (C) =
		\frac{1}{N_c}\,\text{tr}\, \, \mathcal{P} \,\text{exp}\left[i g \oint_{C} \Dif z^\mu A_\mu (z) \right]
\end{equation}
Thus each sub-loop is Lorentz and Dirac invariant, but only together they are evaluated in the ground state to form a $n$-th order Wilson loop. Note that a $\mathcal{W}_1$ loop coincides with our original definition in eq. \eqref{eq:wloop}.
Treating these $n$-th order Wilson loops as elementary objects in loop space, we note that all gauge kinematics are encoded in a $\mathcal{W}_1$ loop. On the other hand, all gauge dynamics are governed by a set of geometrical evolution equations, the Makeenko-Migdal equations \cite{MMeqs}:
\begin{equation}
	\partial^\nu \frac{\delta}{\delta \sigma_{\mu\nu}(x)} \mathcal{W}_1(C) =
		g^2 N_c \oint_C \Dif z^\mu \delta^{(4)}\left(x-z\right) \mathcal{W}_2(C_{xz} \, C_{zx})
\end{equation}
Here a path $\mathcal{C}$ gets deformed by taking two opposite points $x$ and $z$, and bringing them infinitesimally close, such that we separate a newly-formed closed contour from the original one. In other words, we deform the contour $\mathcal{C}$ into two closed contours $\mathcal{C}_{xz}$ and $\mathcal{C}_{zx}$ that are still connected in one point.

Of special importance are the two geometrical operations we introduced in the Makeenko-Migdal equations, namely the path derivative $\partial_\mu$ and the area derivative $\frac{\delta}{\delta\sigma_{\mu\nu}(x)}$ \cite{MMeqs}:
\begin{align}
	\partial_\mu \Phi(C) &=
		\lim_{\left|\delta x_\mu\right|\rightarrow 0}
		\frac{\Phi(\delta x_\mu^{-1} C \, \delta x_\mu) - \Phi(C)}{\left|\delta x_\mu\right|}\\
	\frac{\delta}{\delta\sigma_{\mu\nu}(x)} \Phi(C) &=
		\lim_{\left|\delta \sigma_{\mu\nu}(x)\right|\rightarrow 0}
		\frac{\Phi(C \, \delta C) - \Phi(C)}{\left|\delta \sigma_{\mu\nu}(x)\right|}
\end{align}
The path derivative resembles most our standard notion of a derivative: it measures the variation of the contour while keeping the area constant. On the other hand, the area derivative is the most intuitive interpretation of a geometric derivative: it quantifies the variation of a contour by comparing the original contour $\mathcal{C}$ with a new contour containing small (non area-conserving) deformations $\delta\mathcal{C}$.

Although the Makeenko-Migdal equations provide an elegant method to describe the evolution of a generalised Wilson loop solely in function of its path, they have their limitations. For starters, they are not closed since the evolution of $\mathcal{W}_1$ depends on $\mathcal{W}_2$. Formally, this limitation is superfluous in the large $N_c$ limit since then we can make use of the 't Hooft factorisation property $\mathcal{W}_2 (C_1,C_2)\approx \mathcal{W}_1(C_1)\mathcal{W}_1(C_2)$ \cite{MMeqs}, making the MM equations closed. The remaining limitations of the MM equations are more severe. For one, the evolution equations are derived by applying the Schwinger-Dyson methodology on the Mandelstam formula
\begin{equation}
	\frac{\delta}{\delta \sigma_{\mu\nu} (x)} \Phi(C) =
		i g \,\text{tr}\, \! \left\{ F^{\mu\nu} \Phi(C_x) \right\}
\end{equation}
and using the Stokes' theorem. These might, as well as the area derivative, not be well-defined for all types of paths. In particular, all contours containing one or more cusps (these are non-smooth obstructions, often externally driven) might induce some problematic behaviour, as it is (at least) not straightforward to define \emph{continuous} area differentiation inside a cusp, nor it is to continuously deform a contour in a general topology \cite{ourpaper}. This is somewhat bothersome, as most interesting dynamics lies in contours with cusps.

\section{Evolution of rectangular Wilson loops}
\begin{figure}[h!]
	\label{fig:rect}
	\centering
	\begin{tikzpicture}[scale=0.9]
		\draw[double,double distance=1.5pt] (0,0) -- (1,2) -- (3,2) -- (2,0) -- cycle;
		\path[very thick, postaction={decorate}, decoration={markings,mark=between positions 0.15 and 1 step .25 with {\arrow{>}}}] (0,0) -- (1,2) -- (3,2) -- (2,0) -- cycle;
		\node at (-0.25,-0.25) {$x_1$};
		\node at (0.75,2.25) {$x_2$};
		\node at (3.25,2.25) {$x_3$};
		\node at (2.25,-0.25) {$x_4$};
		\node at (0.15,1) {$v_1$};
		\node at (1.9,2.35) {$v_2$};
		\node at (2.9,1) {$v_3$};
		\node at (1.1,-0.35) {$v_4$};
		\filldraw (0,0) circle(2pt);
		\filldraw (1,2) circle(2pt);
		\filldraw (3,2) circle(2pt);
		\filldraw (2,0) circle(2pt);
	\end{tikzpicture}
	\caption{Parametrisation of a rectangular Wilson loop in coordinate space.}
\end{figure}
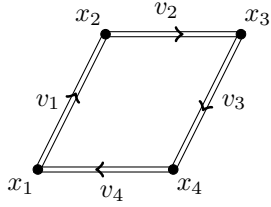
The question at hand is whether it is possible to fully describe the evolution of a general Wilson loop in a geometric but mathematically satisfying way (\emph{i.e.} avoiding non-rigorously defined methods like the Mandelstam formula). We are quite confident that it is highly unprobable to achieve this in a general way. However, one could classify different contours based on their geometrical structure, and treat the evolution on a case-by-case basis. In this paper, we will investigate a rectangular contour with light-like segments on the null-plane, as depicted in fig. (\ref{fig:rect}). Physically, it could represent the soft part of a 4-gluon scattering diagram. It is however more instructive to treat this loop as is; an elementary object in loop space. The final aim is to relate its evolution expressed in coordinate space to a geometric evolution. This is inspired on previous work by \cite{SYM}, where a duality is established between a Wilson loop built from $N$ light-like segments in loop space, and the $N$-gluon scattering amplitude in Super Yang-Mills theories.

The segment lengths
\begin{equation}
	v_i = x_i - x_{i+1}
\end{equation}
are expected to be related to the external momenta (from which the cusps dynamically emerge) and will from now on be treated as such. They are light-like ($v_i^2 = 0$) and on the null-plane ($\bvec{v}_\perp=0$). To investigate its singularity structure, we evaluate the loop \eqref{eq:wloop} at one loop level in coordinate space \cite{wexpansion}:
\begin{equation}
	\mathcal{W}_{\text{L.O.}} =
		1 -\frac{\alpha_s C_F}{\pi} \left(2\pi \mu^2 \right)^\epsilon \Gamma (1-\epsilon) \left[
			\frac{1}{\epsilon^2} \left(-\frac{s}{2} \right)^\epsilon + \frac{1}{\epsilon^2} \left(-\frac{t}{2} \right)^\epsilon
			-\frac{1}{2}\ln^2\frac{s}{t}
		\right]
\end{equation}
where $s$ and $t$ are the Mandelstam energy/rapidity variables (note the positive sign in $t$):
\begin{equation}
	s = \left( v_1 + v_2\right)^2 \qquad\qquad t = \left( v_2 + v_3\right)^2
\end{equation}
Note the $1/\epsilon^2$ poles, which are the overlapping divergencies that stem from the light-like behaviour of the contour segments. The fact that they appear already at leading order renders this kind of Wilson loop non-renormalisable (at least not using the standard $R$-operation). The most straightforward way to manage them is by deriving an evolution equation for the loop. This is done by double differentiation (after rescaling $\bar{s} = \pi e^{\gamma_E} \mu^2 s$):
\begin{equation}\label{eq:evolution}
	\frac{\Dif}{\Dif \ln \mu}\frac{\Dif}{\Dif \ln s} \mathcal{W}_{\text{L.O}} =
		-2\frac{\alpha_s C_F}{\pi} = -2 \Gamma_{\text{cusp}}
\end{equation}
where we recognise the cusp anomalous dimension in the light-cone limit from \eqref{eq:cad}. Thus, as anticipated earlier, the only contribution to the evolution equations stems from the overlapping divergencies. Their concurrent appearance in and similarity to the on-LC TMD case and in the case of an on-LC rectangular Wilson loop again hints to the existence of a duality between both.

\section{Geometrical behaviour}
\begin{figure}[h!]
	\centering
	\begin{tikzpicture}
		\filldraw[fill=black!40] (1,1) -- (1.13,1.26) --  (3.13, 1.26) -- (3,1) -- cycle;
		\filldraw[fill=black!40] (0.65,1) -- (1,1) --  (0, -1) -- (-0.35,-1) -- cycle;
		\begin{scope}[shift={(0,-1)}]
			\draw[double,double distance=1.5pt] (0,0) -- (1,2) -- (3,2) -- (2,0) -- cycle;
			\path[very thick, postaction={decorate}, decoration={markings,mark=between positions 0.15 and 1 step .25 with {\arrow{>}}}] (0,0) -- (1,2) -- (3,2) -- (2,0) -- cycle;
		\end{scope}
		\node[anchor=base] (up) at (2.5,1.6) {$\delta\sigma^{+-}$};
		\node at (1.5,1.15){}
			edge[->, bend left] (up);
		\node[anchor=base] (right) at (-0.4,0.2) {$\;\;\delta\sigma^{-+}$};
		\node at (-0.1,-0.65){}
			edge[->, bend left] (right);
		\node at (1.6 5,0.75) {$v^+$};
		\node at (0.8,-0.25) {$v^-$};
	\end{tikzpicture}
	\caption{Angle-conserving deformations of a light-like rectangular loop on the null-plane.}
	\label{fig:angdefor}
\end{figure}
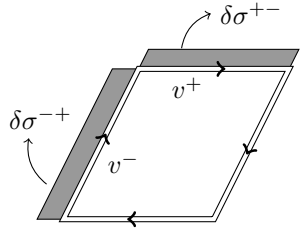
Let start with an attempt to define the area derivative in a more rigorous way. If we don't touch the cusp, we should be save, so we'll only consider contour deformations that conserve the cusp angle, see fig. (\ref{fig:angdefor}). As the cusp angle tends to infinity (because the segments are on-LC), these deformations are sufficient to describe general area variations for this class of contours, rendering the area differentials well-defined \cite{ourpaper}.
This gives
\begin{equation}
\delta\sigma^{+-} =
	\oint \Dif x^- x^+ = v^+\delta v^- \qquad
\delta\sigma^{-+} =
	\oint \Dif x^+ x^- = v^-\delta v^+
 \end{equation}
Next we introduce the area variable $\Sigma$:
\begin{equation}
	\Sigma \equiv 
		v^-\cdot v^+ = \frac{1}{2} s \qquad\qquad
	\frac{\delta}{\delta\ln\Sigma} = 
		\sigma_{\mu\nu}\frac{\delta}{\delta\sigma_{\mu\nu}}
\end{equation}
Replacing $s$ by $\Sigma$ in equation (\ref{eq:evolution}) gives $-4\Gamma_{\text{cusp}}$, in other words, we get one cusp anomalous dimension per cusp in the contour.
Motivated by this, we conjecture a general evolution equation for light-like polygon Wilson loops on the null-plane:
\begin{equation}\label{eq:result}
\frac{\Dif}{\Dif\ln\mu} \left[\sigma_{\mu\nu}\frac{\delta}{\delta \sigma_{\mu\nu}} \ln \mathcal{W} \right] =
						-\sum_i \Gamma_{\text{cusp}}
\end{equation}
Note that this equation is in perfect agreement with the non-Abelian exponentiation of the regularised Wilson loops:
\begin{equation}
	\mathcal{W} =
		\text{exp}\left[ \sum_{k=1} \alpha_s^k C_k \left(\mathcal{W}\right) F_k \left(\mathcal{W}\right) \right]
\end{equation}
where $C_k\sim C_F N_c^{k-1}$ and the summation goes over all `webs' $F_k$, see \cite{webs}.

\section{Relation to TMDs}
Besides for light-like rectangular Wilson loops, equation (\ref{eq:result}) is expected to be valid for light-like TMDs, as they posses the same singularity structure. The area variable then gets replaced by the rapidity variable. This gives
\begin{equation}
	\frac{\Dif}{\Dif \ln \mu}\frac{\Dif}{\Dif \ln \theta} f(x,\bvec{k}_{\perp}) = 2\Gamma_{\text{cusp}}
\end{equation}
The minus disappeared because $\theta\sim \Sigma^{-1}$ ($\theta = \eta / (p\cdot v^-)$ and $p\sim v^+$, so $\theta \sim (v^+v^-)^{-1}$), and there is a factor 2 since we haven two (hidden) cusps. Note that this result is very similar to the Collins-Soper evolution equations for off-LC TMDs. 

We can use the derived formula to get evolution equations for other similar objects, like the $\Pi$-shape Wilson (semi-)loop with one of the segments lying on the light-cone and two semi-infinite off-light-cone sides, see \cite{ourpaper2} for a more detailed analysis.

%%%%%%%%%%%%%%%%%%%%%%%%%%%%%%%%%%%%%%%%%%%%%%%%
%%							BACKMATTER									%%
%%%%%%%%%%%%%%%%%%%%%%%%%%%%%%%%%%%%%%%%%%%%%%%%

\acknowledgments
I'd like to thank I.O.~Cherednikov and T.~Mertens for our conjoint research and for their inspiring discussions. I'd also like to thank the organisers of QCD N'12 for their hospitality and for the wonderful atmosphere during the conference.

\end{document}